\documentclass[%
 aip,
 amsmath,amssymb,
 reprint,%
]{revtex4-1}

\usepackage{graphicx}
\usepackage{dcolumn}
\usepackage{bm}

\usepackage[utf8]{inputenc}
\usepackage[T1]{fontenc}
\usepackage{mathptmx}
\usepackage{etoolbox}

\makeatletter
\def\@email#1#2{%
 \endgroup
 \patchcmd{\titleblock@produce}
  {\frontmatter@RRAPformat}
  {\frontmatter@RRAPformat{\produce@RRAP{*#1\href{mailto:#2}{#2}}}\frontmatter@RRAPformat}
  {}{}
}%
\makeatother
\begin{document}

\preprint{AIP/123-QED}

\title[Space-charge compensation of a low-energy He$^{2+}$ beam \\with fluorine-containing gases]{Space-charge compensation of a low-energy He$^{2+}$ beam \\with fluorine-containing gases}
\author{T. Nagatomo}
 \email{nagatomo@riken.jp}
 \affiliation{RIKEN Nishina Center for Accelerator-Based Science, 2-1 Hirosawa, Wako, Saitama 351-0198 Japan}
\author{Y. Morita}%
 \affiliation{RIKEN Nishina Center for Accelerator-Based Science, 2-1 Hirosawa, Wako, Saitama 351-0198 Japan}

\author{O. Kamigaito}
 \affiliation{RIKEN Nishina Center for Accelerator-Based Science, 2-1 Hirosawa, Wako, Saitama 351-0198 Japan}%

\date{\today}

\begin{abstract}
This study investigates the space-charge compensation (SCC) of a low-energy He$^{2+}$ beam by introducing fluorine-containing gases into the low-energy beam transport line (LEBT) of the RIKEN Heavy-Ion Linear Accelerator.
The introduction of SF$_6$, C$_3$F$_8$, and CHF$_3$ improved beam transmission through the LEBT, and the largest improvement was obtained using SF$_6$.
In contrast, the increase in the beam current with Kr was relatively small, and its effect was less pronounced than that of the fluorine-containing gases.
The introduction of fluorine-containing gases increased the rise time of the chopped beam to approximately 10~ms.
Experiments performed without the prebuncher confirmed that the prebuncher was largely responsible for this slow rise.
Furthermore, the substantial difference in the rise time between Kr and the fluorine-containing gases suggests that different charge carriers may contribute to the SCC in these two cases.
Based on the results, a new grid-type beam chopper was installed immediately upstream of the radio-frequency quadrupole linac (RFQ).
This configuration allows short beam pulses to be generated immediately before the RFQ while maintaining a direct-current beam in the LEBT, thereby achieving both an effective SCC and a fast beam-current rise.
In an acceleration test using the new chopper, a 19.3~$\mu$A He$^{2+}$ beam was successfully and stably accelerated to the radioisotope-production beamline at a duty factor of 10\%.
This corresponds to approximately 193~$\mu$A in continuous-wave mode operation, demonstrating that the proposed method is effective for realizing the acceleration of a 200~$\mu$A-class He$^{2+}$ beam.

\end{abstract}

\maketitle
\section{Introduction}
\label{sec:intro}

The RIKEN heavy-ion linac (RILAC) consists of 9 variable-frequency accelerating cavities\cite{bib:rilac, bib:rfq, bib:csm} that accelerate high-intensity heavy-ion beams in the continuous-wave (CW) mode.
Recently, a superconducting linac booster (SRILAC)\cite{bib:she}, consisting of 10 superconducting cavities, was added to RILAC, and the acceleration of He$^{2+}$ beams up to 30 MeV was initiated to produce $^{211}$At for radiopharmaceutical development.
$^{211}$At is produced through the $^{209}$Bi($\alpha$,2n)$^{211}$At reaction. Moreover, it has short half-life of 7.2 h; therefore, a 200~$\mu$A He$^{2+}$ beam is required to achieve a production rate of 5~GBq/h, which is necessary for clinical trials\cite{bib:production}.
We have been developing the beamline\cite{bib:beamline} and target system\cite{bib:211at, bib:target} necessary to achieve the final goal of a 200~$\mu$A He$^{2+}$ beam.

However, the extraction voltage for the He$^{2+}$ beam at RILAC must be set to a low value of 6.7~kV because of the velocity-matching requirement for injection into the radio-frequency quadrupole linac\cite{bib:rfq}.
Consequently, the He$^{2+}$ beam energy in the low-energy beam transport line (LEBT) is 3.3~${\rm keV}/{\rm u}$ ($\beta=2.7\times10^{-3}$).
In practice, only approximately 200~$\mu$A of the He$^{2+}$ beam can be transported to the end of the LEBT even after extensive optimization of the ion-source gas conditions, RF power, and LEBT magnet settings.
The beam transmission from the LEBT exit to the SRILAC exit is approximately 40\%; therefore, the required beam intensity cannot be achieved under these conditions.
We consider the primary cause of this limitation to be the space-charge effect.

The generalized perveance $K$, which characterizes the strength of the space-charge effect, is given by\cite{bib:reiser}
\begin{eqnarray}
\label{eq:perveance}
K=\frac{2I}{I_0 \beta^3\gamma^3},
\end{eqnarray}
where $I$ is the ion-beam current and $I_0$ is the characteristic current, defined in terms of the vacuum permittivity $\epsilon_0$, ion mass $M$, and charge $Q$ as
\[
I_0=4\pi\epsilon_0 c^3 M/Q.
\]
The $\beta^{-3}$ dependence of the generalized perveance indicates that the space-charge effect becomes particularly important in the LEBT.
Indeed, for a 3.3~${\rm keV}/{\rm u}$, 1~mA He$^{2+}$ beam, $K=1.7\times10^{-3}$, indicating that the space-charge effect cannot be neglected.

For positive-ion beams, the electrons generated by the ionization of residual-gas molecules neutralize the beam charge, thereby reducing the space-charge effect.
The mechanism and degree of this space-charge compensation (SCC) have been extensively investigated.
Reidenbach et al.\cite{bib:epac94} investigated the temporal evolution of the space-charge potential using a 10.7~keV, 3.1~mA He$^+$ beam by performing time-resolved measurements of the energies of residual-gas ions.
Jakob et al.\cite{bib:jakob2000time} introduced Kr gas into a 95~keV, 80~mA H$^+$ beam and reported a reduction in emittance owing to SCC by measuring its temporal evolution.
Lu et al.\cite{bib:ipac11} introduced Kr and Ar into a 95~keV, 80~mA H$^+$ beam and observed decreases in the beam size and emittance while measuring the changes in the space-charge potential using extra compensation gas ions.
Beauvais et al.\cite{bib:silhi2000} investigated the temporal evolution of SCC using a 92~keV, 62~mA H$^+$ beam based on measurements with a residual-gas ion analyzer.
More recently, Cosgun et al.\cite{bib:kaeri} investigated a 25~keV, 12--18~mA H$^+$ beam with Kr gas and reported the temporal evolution of the emittance and beam size in detail.
A comprehensive overview of the measurement techniques and evaluation methods for SCC is provided in Ref.~\cite{bib:ibic25}.

Thus, previous studies on SCC have mainly focused on compensation using noble gases, such as Kr and Ar, whereas SCC using fluorine-containing gases has rarely been reported.
However, an increase in the beam current was reported recently for a low-energy ion beam used in semiconductor-manufacturing equipment by introducing fluorine-containing gas C$_3$F$_8$\cite{bib:kuwata2024}.
This suggests that fluorine-containing gases may be effective in improving the transport of low-energy ion beams.

In this study, C$_3$F$_8$, CHF$_3$, and SF$_6$ were introduced into the RILAC LEBT, and their effects on the SCC of the He$^{2+}$ beam were systematically investigated.
Furthermore, a new beam chopper was developed to overcome the increased beam rise time, which presents a practical limitation when applying this method to accelerator operations.
This study reports the SCC effects obtained employing fluorine-containing gases and the acceleration of a high-intensity He$^{2+}$ beam using a newly developed beam chopper.

\section{Experimental Setup}

\subsection{Beamline Layout}

\begin{figure*}[htbp]
   \centering
   \includegraphics[width=145 mm]{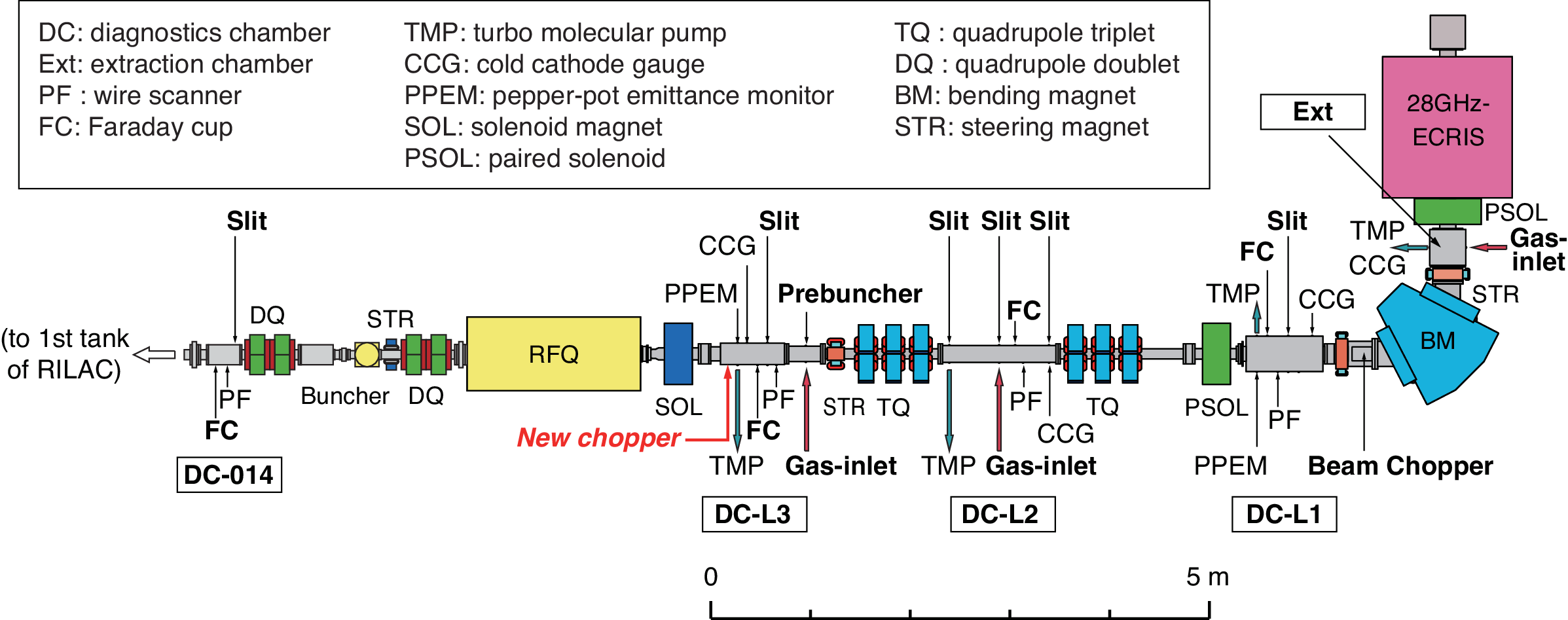}
   \caption{Schematic of the RILAC beamline.
   The LEBT is designed to limit the beam emittance using two TQs and a three-stage slit system\cite{bib:lebt}.
   A deflector-type chopper is installed immediately downstream of the BM, and a grid-type prebuncher is installed in DC-L3 immediately upstream of the RFQ.
   Neutralizing gases are introduced through three diagnostic chambers in the LEBT and evacuated by vacuum pumps.
   The grid-type chopper developed in this study was installed immediately upstream of the RFQ.}
   \label{fig:lebt}
\end{figure*}

Figure~\ref{fig:lebt} shows the schematic layout of the upstream RILAC beamline including the LEBT.
The design concept of the LEBT is described in detail in Ref.~\cite{bib:lebt}.
In addition to the bending magnet (BM), the LEBT comprises two quadrupole triplets (TQs) and three paired solenoids (PSOLs).
Each PSOL consists of two independently excited solenoid coils housed in a common yoke. They are designed to cancel the beam rotation induced by the solenoidal magnetic fields while providing a net focusing effect.
Three diagnostic chambers, DC-L1, DC-L2, and DC-L3, are installed between these magnet systems.
The beam optics are designed to form beam waists at two locations: the slit position at the center of DC-L2 and the prebuncher position in DC-L3. Moreover, the beam phase-space distribution is matched to the RFQ acceptance.
The design acceptance of the LEBT is 300~mm$\cdot$mrad.

The root mean square emittance of the He$^{2+}$ beam extracted at 6.7~kV from the electron-cyclotron resonance ion source (ECRIS) is estimated to be approximately 150~mm$\cdot$mrad from the extraction-side mirror magnetic field ($B_{\rm ext}=1.6~{\rm T}$)\cite{bib:Leitner}.
Furthermore, the beam emittance is limited by the three-stage slit system installed in DC-L2; therefore, the maximum beam transmission from DC-L1 to DC-L3 is designed to be approximately 50~\% in the absence of space-charge effects.

This study investigates the effect of gas injection on the SCC by introducing neutralizing gases through variable leak valves into the three diagnostic chambers, DC-L1, DC-L2, and DC-L3, and the extraction chamber, Ext, at the ion-source exit, as shown in Fig.~\ref{fig:lebt}.
Each diagnostic chamber is equipped with a wire scanner (PF), Faraday cup (FC), turbomolecular pump (TMP), and cold-cathode gauge (CCG).
The pumping speed of the TMP in DC-L1 is 450~L/m, whereas those in DC-L2 and DC-L3 are 220~L/m.
The FCs in DC-L1 and DC-L2 are equipped with permanent-magnet secondary-electron suppressors, whereas the FC in DC-L3 is equipped with an electrostatic suppressor electrode.
Pepper-pot emittance monitors\cite{bib:ppem, bib:optical, bib:PPEM_NIMA} are installed in DC-L1 and DC-L3.
Ext is not equipped with beam-diagnostic devices; however, its vacuum is maintained by a large TMP with a pumping speed of 1100~L/m and is monitored by a CCG.
The base pressure in these vacuum chambers is approximately $10^{-5}$~Pa.

A beam chopper is installed immediately downstream of the BM.
It is a deflector-type chopper comprising parallel-plate electrodes with a gap of 120~mm and length of 200~mm.
A voltage of up to 2~kV is applied between the electrodes to deflect the beam vertically downward.
The rise time to reach the set voltage is 40~ns, and the fall time to reach 0~V is 100~ns.

The temporal variation in the beam current was measured by terminating the Faraday cup with a 1~k$\Omega$ resistor, and the voltage across the resistor was measured using an oscilloscope (DSO-X 4024A, Keysight Technologies, USA).
Thus, 1~V corresponds to a beam current of 1~mA.

Each slit consists of four Ta plates, with two vertical and two horizontal plates that can be moved independently.
Each slit plate is connected to an independent logarithmic amplifier to measure the incident beam current on the plate.
The input impedance of the logarithmic amplifier is 500~k$\Omega$.
The slit plates acquire a positive potential under beam irradiation because of this high input impedance, thereby suppressing secondary-electron emission and enabling an accurate measurement of the beam current.

A prebuncher is installed in the upstream part of DC-L3 to improve the overall acceleration efficiency of the RILAC--SRILAC system\cite{bib:prebuncher}.
The prebuncher consists of two opposing copper-grid electrodes.
The grids have a honeycomb structure, and each has a transmission of 90~\%\cite{bib:mesh}.
A 200~$\Omega$ resistor is connected between the two electrodes, and its impedance is transformed to 50~$\Omega$ using a 1:4 transmission-line transformer.
For direct-current signals, both grid electrodes are at the ground potential.
The three-stage slit system and prebuncher are essential components for stable beam delivery to RILAC.

\subsection{Experimental condition}

A high-intensity $^{4}$He$^{2+}$ beam of approximately 1~mA was produced using the RILAC 28~GHz superconducting ECRIS R28G-K\cite{bib:lebt}.
The R28G-K consists of six superconducting solenoid coils and one superconducting hexapole coil.
The axial mirror magnetic fields were set to $B_{\rm inj.}=2.3~{\rm T}$, $B_{\rm min.}=0.55~{\rm T}$, and $B_{\rm ext.}=1.6~{\rm T}$.
The ECR plasma was heated using 18 and 28~GHz microwaves, and the respective microwave powers were adjusted to maintain stable plasma conditions; the microwave powers were $P$(18~GHz)=500--700~W and $P$(28~GHz)=1.5--2~kW, respectively.
He gas was introduced into the plasma chamber through a variable leak valve and ionized in the ECR plasma.
No supporting gas was used.
An ion-extraction voltage of $V_{\rm ext.}=6.7~{\rm kV}$ was applied to the electrically isolated plasma chamber.
The gap between the plasma chamber and extraction electrode and extraction solenoid-lens current (approximately 90~A) were adjusted to maximize the beam current at DC-L3.

\begin{figure}[htbp]
   \centering
   \includegraphics[width=80 mm]{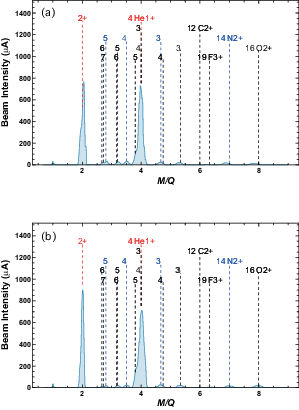}
   \caption{Extracted-beam $M/Q$ spectra (a) without gas injection and (b) with SF$_6$ injection.
   Both spectra were measured using the FC in DC-L1.
   Ion-beam components other than He were negligibly small.}
   \label{fig:spectrum}
\end{figure}

A typical $M/Q$ spectrum is shown in Fig.~\ref{fig:spectrum}(a).
He ions were the dominant components, and only small additional peaks, which were considered to originate from residual gases such as nitrogen and oxygen in the ion source, were observed.

The SCC effect was investigated by introducing gases into Ext, DC-L2, and DC-L3 through variable leak valves to control the gas-flow rates.
The beam currents at DC-L1, DC-L2, and DC-L3 were measured for each condition, together with the temporal variation in the beam current at DC-L3 when the beam was chopped.

The gases used in this study were Kr, C$_3$F$_8$, SF$_6$, and CHF$_3$.
C$_3$F$_8$ and SF$_6$ have large electron-attachment cross sections for low-energy electrons.
Electron attachment to C$_3$F$_8$ occurs mainly at energies of a few electron volts\cite{bib:c3f8}, whereas SF$_6$ exhibits a very strong electron attachment even at lower electron energies\cite{bib:sf6}.
In contrast, significant electron attachment to CHF$_3$ has not been reported\cite{bib:chf3}; therefore, CHF$_3$ was included as a comparison gas.
An example of the $M/Q$ spectrum obtained with SF$_6$ injection is shown in Fig.~\ref{fig:spectrum}(b).

The pressures were measured using the CCGs installed in Ext, DC-L1, DC-L2, and DC-L3, as shown in Fig.~\ref{fig:lebt}.
The CCG readings were calibrated using the average conversion factors provided in the manufacturer's datasheet.
However, for CHF$_3$ and C$_3$F$_8$, the conversion factors were estimated based on Ref.~\cite{bib:nishimura1999} and may exhibit an uncertainty of approximately 20~\%.
The conversion factors used in this study are listed in Table~\ref{tab:ConvFactors}.

\begin{table}[hbt]
   \centering
   \caption{Conversion factor $C$ for the CCG reading for each gas species.
   The actual pressure $P$ was obtained from the CCG reading $P_{\rm obs}$ as
   $P=C\,P_{\rm obs}$.}
   \begin{tabular}{cccccc}
       \hline
        & Air & Kr & CHF$_3$ & C$_3$F$_8$ & SF$_6$ \\
       \hline
       \hline
       $C$ & 1.00 & 0.52 & 0.55 & 0.25 & 0.42 \\
       \hline
   \end{tabular}
   \label{tab:ConvFactors}
\end{table}

Table~\ref{tab:vacuum} summarizes the introduced gas species and pressures measured on each experimental date.
The amount of each introduced gas was adjusted to maximize the beam current at DC-L3.

\begin{table}[hbt]
   \centering
   \caption{Gas species and pressures for each measurement.
   "Ext" denotes the pressure in the extraction chamber Ext, and "L1"--"L3" denote the calibrated pressures in the diagnostic chambers DC-L1--DC-L3, respectively.
   Date denotes the measurement date (YYMMDD).}
   \begin{tabular}{lcrrrrl}
       \hline
       Gas & Date & Ext & L1 & L2 & L3 & \\
       \hline
       \hline
           (None) & 260318 & 4.0 & 2.7 & 1.9 & 0.89 & $\times 10^{-5}$ [Pa] \\
           Kr & 260318 & 42 & 21 & 12 & 15 & $\times 10^{-5}$ [Pa] \\
           CHF$_3$ & 260318 & 209 & 105 & 39 & 20 & $\times 10^{-5}$ [Pa] \\
           \hline
           (None) & 260319 & 5.0 & 3.4 & 1.2 & 0.93 & $\times 10^{-5}$ [Pa] \\
           C$_3$F$_8$ & 260319 & 65 & 35 & 16 & 3 & $\times 10^{-5}$ [Pa] \\
           \hline
           (None) & 260317 & 3.9 & 2.9 & 1.9 & 1.7 & $\times 10^{-5}$ [Pa] \\
           SF$_6$ & 260317 & 39 & 23 & 42 & 16 & $\times 10^{-5}$ [Pa] \\
       \hline
   \end{tabular}
   \label{tab:vacuum}
\end{table}

\section{Results}
\label{sec:results}

Table~\ref{tab:transmission} summarizes the beam currents and beam-transmission efficiencies for each gas.
The measurement dates are also listed because the operating conditions of the ion source varied slightly between days, resulting in small variations in the beam current, even without gas injection.
The LEBT-magnet parameters were fixed at settings that provided the maximum beam current when an approximately 150~$\mu$A He$^{2+}$ beam was accelerated immediately upstream of SRILAC with a C$_3$F$_8$ injection.

\begin{figure}[htbp]
   \centering
   \includegraphics[width=80 mm]{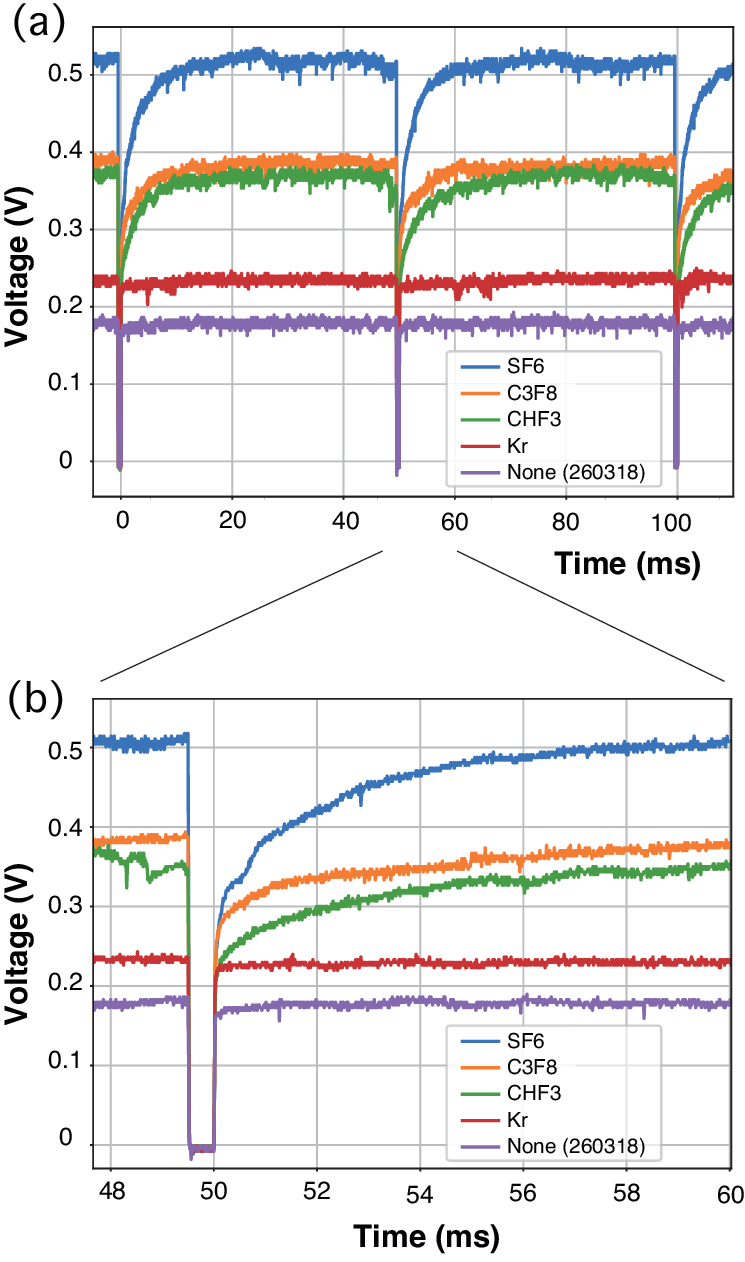}
   \caption{Temporal response of the chopped beam at DC-L3.
   The measurements were performed with a duty factor of 99~\% and a repetition rate of 20~Hz.
   (a) shows the entire waveform over 100~ms, and (b) shows an expanded view of the 12~ms interval around the beam rise.
   The gas species used are indicated in the figure.
   "None" denotes the measurement without gas injection, for which the data obtained on March 18, 2026 (260318) were used (see Table~\ref{tab:vacuum}).}
   \label{fig:buildup}
\end{figure}

Figure~\ref{fig:buildup} shows the temporal response of the chopped beam at DC-L3.
The measurements were performed with a duty factor of 99~\% and repetition rate of 20~Hz.
The upper panel shows the entire waveform over 100~ms, whereas the lower panel shows an expanded view of the 12~ms interval around the beam rise.

Two notable features can be identified in Table~\ref{tab:transmission} and Fig.~\ref{fig:buildup}.

First, the introduction of fluorine-containing gases substantially increased the beam current $I_1$ at DC-L1 and improved the beam-transmission efficiency.
Consequently, the beam current $I_3$ at DC-L3 also increased.
The largest improvement was obtained for SF$_6$.
Notably, no adjustments were made to the ion source that could account for such large changes in the beam current; thus, the observed increases were induced by gas injection.
The beam current also increased slightly with Kr injection; however, the effect was smaller than that obtained with fluorine-containing gases.

Second, the rise time of the chopped beam at DC-L3 strongly depended on the gas species.
Without gas injection and with Kr injection, the rise time was very short.
In contrast, with the SF$_6$, C$_3$F$_8$, and CHF$_3$ injections, the rise time increased to approximately 10~ms.

\begin{table}[hbt]
   \centering
   \caption{Beam currents and beam-transmission efficiencies for each gas.
   $I_1$, $I_2$, and $I_3$ are the beam currents measured using the FCs at DC-L1, DC-L2, and DC-L3, respectively.
   The transmission efficiency is defined as $\epsilon_{ij}\equiv I_j/I_i~(i<j)$.
   The pressures are listed in Table~\ref{tab:vacuum}.
   Date denotes the measurement date (YYMMDD).}
   \begin{tabular}{lccccccc}
       \hline
       Gas & Date & $I_1$[$\mu$A] & $I_2$[$\mu$A] & $I_3$[$\mu$A] & $\epsilon_{12}$ & $\epsilon_{23}$ & $\epsilon_{13}$  \\
       \hline
       \hline
           (None) & 260318 & 500 & 437 & 185 & 0.87 & 0.42 & 0.37\\
           Kr & 260318 & 654 & 517 & 234 & 0.79 & 0.45 & 0.36\\
           CHF$_3$ & 260318 & 852 & 681 & 373 & 0.80 & 0.55 & 0.44\\
           \hline
           (None) & 260319 & 574 & 505 & 223 & 0.88 & 0.44 & 0.39\\
           C$_3$F$_8$ & 260319 & 867 & 672 & 382 & 0.78 & 0.57 & 0.44\\
           \hline
           (None) & 260317 & 644 & 525 & 286 & 0.82 & 0.54 & 0.44\\
           SF$_6$ & 260317 & 1007 & 924 & 550 & 0.92 & 0.60 & 0.55\\
       \hline
   \end{tabular}
   \label{tab:transmission}
\end{table}

\section{Discussion}
\label{sec:discussion}

As shown in Section~\ref{sec:results}, the introduction of fluorine-containing gases improved beam transmission through the LEBT and increased the rise time of the chopped beam to approximately 10~ms.
These results strongly suggest that fluorine-containing gases enhance SCC in the LEBT.

\begin{figure}[htbp]
   \centering
   \includegraphics[width=80 mm]{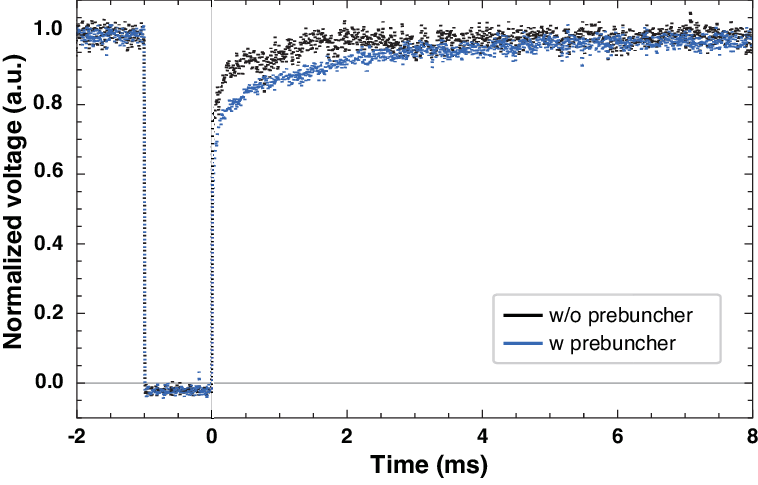}
   \caption{Comparison of the beam rise time with and without the prebuncher.
   The black curve shows the result with the prebuncher removed, whereas the blue curve shows the result with the prebuncher installed.
   For comparison, both curves are normalized to the steady-state beam current.}
   \label{fig:bump_NoSlit}
\end{figure}

The prebuncher, which can intercept part of the beam in the LEBT, was removed, and its effect on the rise time was examined to investigate the origin of the slow rise of the chopped beam.
The results are presented in Fig.~\ref{fig:bump_NoSlit}.
The rise time shortened when the prebuncher was removed, indicating that the prebuncher plays a significant role in the slow rise observed with fluorine-containing gases.

For positive-ion beams, the electrons generated by collisions between the beam and residual-gas molecules can act as charge carriers for SCC.
When fluorine-containing gases are introduced, some of the generated electrons can be converted into negative ions via electron attachment.
In addition, the slit plates and grid electrodes of the prebuncher may absorb electrons or negative ions that contribute to SCC.

Electrons and negative ions differ significantly in mass; therefore, their transport properties are also expected to differ significantly.
Electrons have high mobility; therefore, they can be rapidly lost from the vicinity of the beam. However, negative ions have a considerably lower mobility and may remain near the beam for a longer period.
Consequently, negative ions may gradually accumulate under continuous beam irradiation, resulting in a higher degree of SCC.
This accumulation process requires a finite time; thus, SCC would not be fully established immediately after the beam is turned on. Therefore, the beam transmission would be lower than that in the steady state.
The rise time of approximately 10~ms observed in the present study is consistent with this perspective.

Notably, the charge carriers responsible for SCC were not directly measured in the present study; therefore, their identities could not be conclusively determined.
However, the substantial difference in the rise times of Kr and the fluorine-containing gases, together with the pronounced change in the temporal response caused by removing the prebuncher, suggests that different charge carriers may contribute to SCC in the two cases.

Beam tuning is often initiated in RILAC using a chopper at a repetition rate of 500~Hz and duty factor as low as 0.4~\% to protect the accelerator components.
However, for such short beam pulses, SCC with fluorine-containing gases may not have sufficient time to be fully established.
Consequently, the accelerator parameters optimized under these conditions may no longer be optimal when the duty factor is increased for long-pulse or CW operations.

Therefore, for the practical use of SCC with fluorine-containing gases, a direct-current beam should be transported continuously through the LEBT so that a steady-state SCC condition can be maintained around the prebuncher.
Therefore, an effective configuration is to install an additional beam chopper immediately upstream of the RFQ and generate short beam pulses only after the beam passes through the LEBT.
These results indicate that this beamline configuration is suitable for utilizing SCC with fluorine-containing gases.

\section{New Beam Chopper and Present Status}
\label{sec:newchopper}

\begin{figure}[htbp]
   \centering
   \includegraphics[width=80 mm]{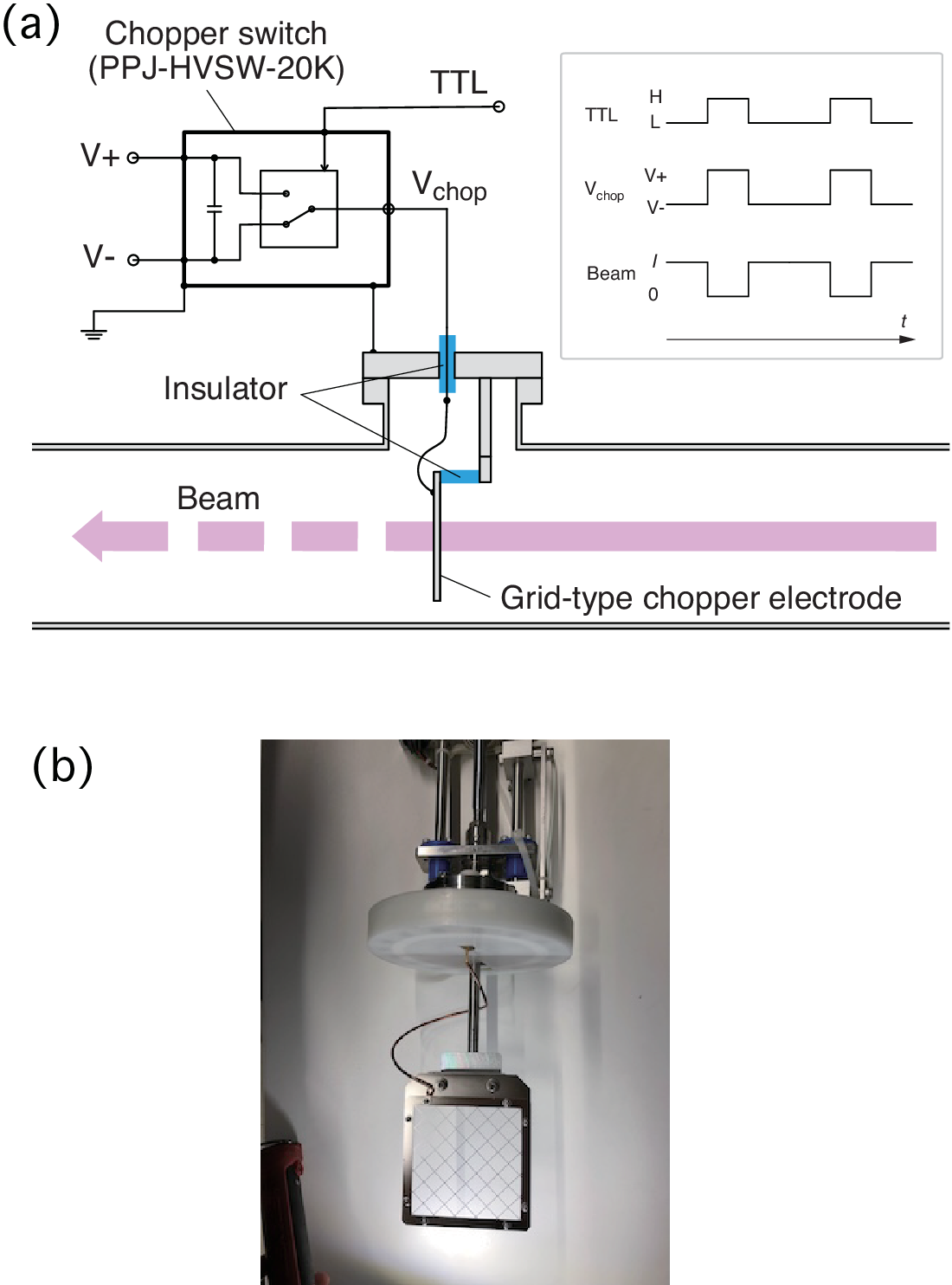}
   \caption{(a) Schematic and (b) photograph of the new beam chopper.
   The timing sequence of the driving voltage is also shown in the upper-right corner of (a).
   In this study, the chopper was operated with $V_{+}=8.2~\mathrm{kV}$ and $V_{-}=-150~\mathrm{V}$.}
   \label{fig:L3_Chopper}
\end{figure}

Based on the discussion in the preceding section, we developed a new beam chopper and installed it in DC-L3 immediately upstream of the RFQ.
The available installation space in DC-L3 was limited; therefore, a stainless-steel grid electrode with an 80~mm $\times$ 80~mm aperture and open-area ratio of 94~\% was adopted instead of the conventional parallel-plate configuration, as shown in Fig.~\ref{fig:L3_Chopper}.
To block the beam, a voltage of 8.2~kV was applied to the electrode during the beam-off period, and the voltage was rapidly switched to $-150$~V using a high-speed switch during the beam-on period.
The voltage-switching time measured using a high-voltage probe was less than 1~$\mu{\rm s}$.

The new chopper was installed immediately upstream of the RFQ; therefore, a direct-current beam can continuously pass through the slits and prebuncher in the LEBT.
Therefore, this configuration is expected to suppress the increase in the beam rise time observed with the injection of fluorine-containing gas.

The beam rise times were compared between two configurations to verify this effect: one using the new chopper immediately upstream of the RFQ and the other using the conventional chopper upstream of DC-L1.
The new chopper is located in the downstream part of DC-L3; therefore, the beam rise time was measured at DC-014 after acceleration through the RFQ.

\begin{figure}[htbp]
   \centering
   \includegraphics[width=80 mm]{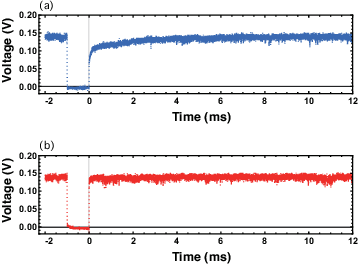}
   \caption{Beam rise at DC-014 for beam chopping with a duty factor of 99~\% and repetition rate of 10~Hz.
   (a) shows the result obtained using the conventional chopper upstream of DC-L1, and
   (b) shows the result obtained using the new chopper immediately upstream of the RFQ.}
   \label{fig:A014_10Hz}
\end{figure}

\begin{figure}[htbp]
   \centering
   \includegraphics[width=80 mm]{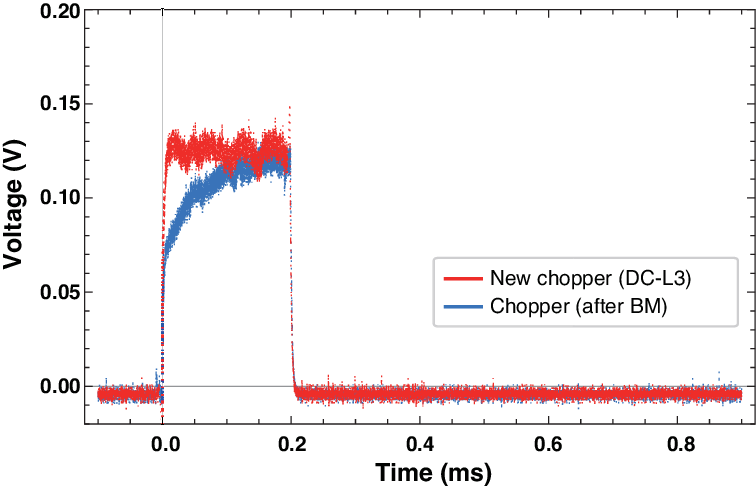}
   \caption{Beam rise at DC-014 for beam chopping with a repetition rate of 500~Hz and duty factor of 10~\% (pulse width of 0.2~ms).
   The blue curve shows the result obtained using the conventional chopper upstream of DC-L1, whereas the red curve shows the result obtained using the new chopper immediately upstream of the RFQ.}
   \label{fig:A014_500Hz}
\end{figure}

Figure~\ref{fig:A014_10Hz} shows the results obtained with a duty factor of 99~\% and repetition rate of 10~Hz, which were used to compare the overall beam-rise behavior.
In contrast, Fig.~\ref{fig:A014_500Hz} shows the results obtained with a repetition rate of 500~Hz and duty factor of 10~\%, which are close to the conditions used in actual accelerator operation.
Under both conditions, the installation of the new chopper immediately upstream of the RFQ substantially reduced the beam rise time.

Acceleration tests including SRILAC are currently being performed using this new chopper.
A 19.3~$\mu$A He$^{2+}$ beam was successfully and stably accelerated toward the radioisotope (RI) production beamline with a duty factor of 10~\% (repetition rate of 500~Hz)\cite{bib:ms2026}.
This corresponds to approximately 193~$\mu$A in CW operation, which is close to the final target beam current of 200~$\mu$A.

\section{Conclusion}

In this study, the effect of fluorine-containing gases on the SCC of a low-energy He$^{2+}$ beam was investigated in the LEBT of RILAC.
The introduction of SF$_6$, C$_3$F$_8$, and CHF$_3$ improved the beam transmission through the LEBT, and the largest improvement was obtained using SF$_6$.
In contrast, the increase in the beam current with Kr was relatively small, and the improvement was less pronounced than that obtained with fluorine-containing gases.

The introduction of fluorine-containing gases increased the rise time of the chopped beam to approximately 10~ms.
Experiments performed with the prebuncher removed demonstrated that the prebuncher plays a significant role in this slow rise.
Furthermore, the substantial difference in the rise time between Kr and the fluorine-containing gases suggests that different charge carriers may contribute to SCC in the two cases.

Based on these findings, a new grid-type beam chopper was developed for installation immediately upstream of the RFQ.
This configuration enabled short beam pulses to be generated immediately before the RFQ while maintaining a direct-current beam through the LEBT, thereby achieving an effective SCC with fluorine-containing gases and a short beam rise time.

In an acceleration test using the new chopper, a 19.3~$\mu$A He$^{2+}$ beam was successfully and stably accelerated to the RI production beamline with a duty factor of 10~\%.
This corresponds to approximately 193~$\mu$A in CW operation and demonstrates that the present method provides an effective approach for realizing a 200~$\mu$A-class He$^{2+}$ beam.

Future work will focus on applying this approach to ion species other than He$^{2+}$.

\section*{ACKNOWLEDGMENTS}

We are grateful to all members of the RILAC operations staff for their assistance in carrying out the extensive measurements required for this study.
We also thank the members of the Ion Source and Injector Team for their valuable discussions.
We thank Dr. Takemura of Nissin Ion Equipment Co., Ltd. for bringing Ref.~\cite{bib:kuwata2024} to our attention.
This research was supported by commissioned research funding from the F-REI (JPFR 25040201 and JPFR 26040201).
This study was partially supported by the Japan Society for the Promotion of Science (JSPS) KAKENHI (Grant No. JP26K15479).
A Japanese patent application related to this work has been filed (Japanese Patent Application No. 2026-096784).
We acknowledge the use of ChatGPT for language assistance in improving the clarity and readability of the manuscript.
Finally, we would like to thank Editage (www.editage.jp) for English language editing.

\section*{Data Availability}

The data that support the findings of this study are available
from the corresponding author upon reasonable request.

\bibliography{aipsamp}

\end{document}